\documentclass[%
 reprint,
superscriptaddress,
 amsmath,amssymb,
 aps,
 prl,
]{revtex4-1}

\usepackage{graphicx}% Include figure files
\usepackage{dcolumn}% Align table columns on decimal point
\usepackage{bm}% bold math
\begin{document}

\preprint{APS/123-QED}

\title{Inelastic light scattering measurements of a pressure-induced quantum liquid in KCuF$_3$}

\author{S. Yuan}
\affiliation{%
Department of Physics and Frederick Seitz Materials Research Laboratory, University of Illinois, Urbana, Illinois 61801, USA
}%

\author{M. Kim}
\affiliation{%
Department of Physics and Frederick Seitz Materials Research Laboratory, University of Illinois, Urbana, Illinois 61801, USA
}%

\author{J. T. Seeley}
\affiliation{%
Department of Physics and Frederick Seitz Materials Research Laboratory, University of Illinois, Urbana, Illinois 61801, USA
}%

\author{J.C.T. Lee}
\affiliation{%
Department of Physics and Frederick Seitz Materials Research Laboratory, University of Illinois, Urbana, Illinois 61801, USA
}%

\author{ S. Lal}
\affiliation{%
Department of Physics and Frederick Seitz Materials Research Laboratory, University of Illinois, Urbana, Illinois 61801, USA
}%
\affiliation{Department of Physical Sciences, IISER-Kolkata, Mohanpur campus, West Bengal - 741252, India}

\author{P. Abbamonte}
\affiliation{%
Department of Physics and Frederick Seitz Materials Research Laboratory, University of Illinois, Urbana, Illinois 61801, USA
}%

\author{S.L. Cooper}
\affiliation{%
Department of Physics and Frederick Seitz Materials Research Laboratory, University of Illinois, Urbana, Illinois 61801, USA
}%

\date{\today}% It is always \today, today,

\begin{abstract}
Pressure-dependent, low temperature inelastic light (Raman) scattering 
measurements of KCuF$_3$ show that applied pressure above $P^{*} \sim$ 7 kbar suppresses a previously observed structural phase transition temperature to zero temperature in KCuF$_3$, resulting in the development of a 
$\omega\sim$ 0 fluctuational (quasielastic) 
response near $T \sim$ 0 K.  This pressure-induced fluctuational response --- which we associate with slow fluctuations of the CuF$_6$ octahedral orientation --- is temperature
 independent and exhibits a characteristic fluctuation rate that is much larger than the 
temperature, consistent with quantum fluctuations of the CuF$_6$ octahedra. A model of pseudospin-phonon 
coupling provides a qualitative description of both the temperature- and pressure-dependent evolution of 
the Raman spectra of KCuF$_3$.

\end{abstract}

%\pacs{42.50, -p}% PACS, the Physics and Astronomy
\maketitle
\thispagestyle{empty}
\pagestyle{empty}

%\tableofcontents

Frustrated magnetic systems in which conventional magnetic order is suppressed down to $T$ = 0 K are currently of great 
interest, because these systems can exhibit exotic phenomena, e.g., off-diagonal long 
range order,[1] and novel ``liquid-like'' ground states --- such as orbital [2] and spin liquids [3] --- that quantum 
mechanically fluctuate even at $T$ = 0 K. Unfortunately, there are only a few examples of real materals in which such fluctuating ground states have been reported.[2,3]

In this paper, we report the first spectroscopic evidence for a pressure-tuned quantum melting transition in KCuF$_3$ between a 
static structural phase to a phase in which fluctuations persist even at $T \sim$ 0 K. 
While often considered a model system for orbital-ordering behavior,[4] the 3$d^9$ perovskite KCuF$_3$ is known to exhibit a number of unusual properties that are still not well understood,[5--16] including 
a highly anisotropic exchange coupling $\left( J_c/J_a\sim -100\right) $ [5] that results in 1D antiferromagnetic 
Heisenberg spin dynamics above 40 K,[6--8] and a large disparity between 
the orbital ordering temperature  ($ T_{oo}\sim 800$ K [9]) and the  N$\acute{e}$el ordering 
temperature  ($ T_{N}\sim 40$ K [5,8]) that cannot be explained by conventional 
superexchange models.[10] Pressure-dependent, low temperature inelastic light (Raman) scattering measurements reported here show 
that applied pressure above $P^{*} \sim$ 7 kbar suppresses a previously observed structural phase transition temperature [15, 16] in KCuF$_3$ down to the lowest temperatures measured ($T$ = 3 K), resulting in the 
development of a quasielastic response that is indicative of fluctuational dynamics near $T \sim$ 0 K. This pressure-induced fluctuational response --- which we associate with slow fluctuations of the CuF$_6$ octahedra between discrete orientations --- is temperature independent and exhibits a characteristic fluctuation rate that is much 
larger than the temperature, similar to the behavior observed in ``quantum paraelectric'' phases in SrTiO$_3$ and KNaO$_3$.[1] A model of pseudospin-phonon coupling [17] --- where the pseudospin is identified with different CuF$_6$ octahedral rotational configurations --- is qualitatively consistent with our results on KCuF$_3$ and shows that KCuF$_3$ can be systematically tuned with pressure and temperature between the  characteristic ``soft-phonon'' and ``diffusive mode'' regimes predicted for strongly pseudospin-phonon coupled systems.[17]  

Single crystal samples of KCuF$_3$ were grown by an aqueous solution
 precipitation method described previously.[18] Samples were characterized with
 magnetic susceptibility and X-ray diffraction measurements, and the results obtained
 are in good agreement with previous results.[6,7,19] Low temperature, pressure-dependent 
Raman scattering measurements --- using liquid argon as the quasihydrostatic 
pressure medium --- were performed using the 6471 $\AA$ line from a krypton laser
 and a SiC- or diamond-anvil cell that fits in a flow-through helium cryostat, allowing simultaneous  \textsl{in situ} control of the sample temperature ($T$ $\textgreater$ 3 K) and pressure ($P$ $\textless$ 100 kbar).
\begin{figure}
\centering
\includegraphics[totalheight=3.2 in,width=3.4 in,  scale=0.5, angle=0]{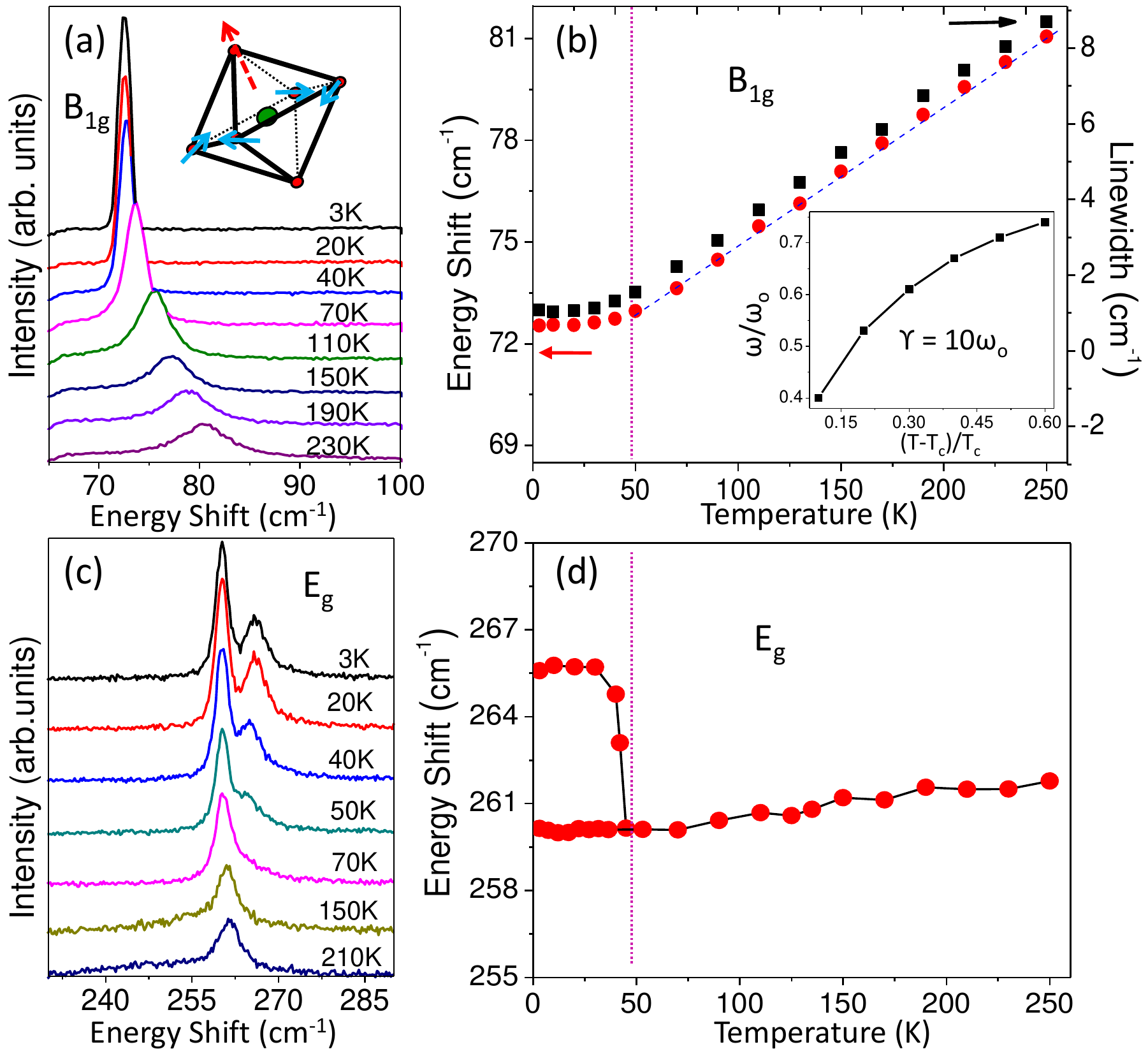}
\caption{(a) Temperature dependence of the $B_{1g}$ symmetry phonon mode in KCuF$_3$. All spectra have the same y-axis scale and have been offset in the y-axis direction for clarity. Inset shows $B_{1g}$ phonon normal mode vibration of F$^{-}$ ions (blue arrows), and dashed red arrow depicts octahedral orientation (pseudospin).  (b) Summary of the temperature dependence of the peak energy (circles) and linewidth (squares) of the $B_{1g}$  phonon mode. The inset shows the calculated temperature dependence of 
the normalized peak frequencies, $\omega/ \omega_o$, using Eq. 1 for the case $\gamma = 10\omega_{o}$, from [17].  (c) Temperature dependence of the $E_g$ symmetry 
phonon mode in KCuF$_3$.  All spectra have the same y-axis scale and have been offset in the y-axis direction for clarity. (d) Summary of the temperature dependence of the 
peak energy of the $E_g$ phonon mode, showing a splitting of the mode at the tetragonal-to-orthorhombic structural transition at $T$ = 50 K.}
\label{fig:test}
\end{figure}

Fig. 1 summarizes the temperature-dependence ($P$ = 0) of some of the key phonon modes in KCuF$_3$,[16,20] showing 
evidence for a structural phase transition in KCuF$_3$ at $T$ = 50 K.  In particular, Figs. 1 (a) and (b) show that the $B_{1g}$-symmetry 
phonon near 72 cm$^{-1}$ exhibits a roughly 10-fold decrease in linewidth (FWHM) and a 20$\%$ decrease in 
energy (``softening'') with decreasing temperature (Fig. 1), consistent with previous evidence for thermally 
driven structural fluctuations that persist over a broad range of temperatures 
between $T_{N}$ (=40 K) and 300 K.[11--14,16] Fig. 1(b) also shows that the $B_{1g}$ phonon frequency stabilizes at 
temperatures below $\sim$ 50 K, concomitant with a splitting of the doubly degenerate 260 cm$^{-1} E_{g}$ mode into two 
singly degenerate modes at 260 cm$^{-1}$ and 265 cm$^{-1}$ (Figs 1 (c) and (d)); this result provides evidence that the thermally driven structural fluctuations in KCuF$_3$ are arrested by a tetragonal-orthorhombic structural distortion that locks the CuF$_6$ octahedral tilt orientations into a static, `glassy' configuration at $T$ = 50 K. [16]

Evidence that CuF$_6$ octahedral fluctuations in KCuF$_3$ extend down to very low temperatures ($\sim$ 50 K) --- and are interrupted only by a tetragonal-to-orthorhombic distortion --- suggests that KCuF$_3$ is close to a quantum critical point at which the fluctuational regime extends down to $T$ = 0 K. Hydrostatic pressure has been shown to reduce octahedral 
distortions in perovskite materials such as (La,Ba)$_2$CuO$_4$,[21] Ca$_2$RuO$_4$,[22] Ca$_3$Ru$_2$O$_7$,[23] and LaMnO$_3$;[24] therefore, pressure tuning offers a means of suppressing to $T$ = 0 K the low-temperature tetragonal-to-orthorhombic distortion in KCuF$_3$ that locks in CuF$_6$ octahedral rotations below $T$ = 50 K (and $P$ = 0). For this reason, we performed low-temperature, pressure-dependent Raman scattering measurements on KCuF$_3$ in an effort to induce and study ``quantum melting'' between $T \sim$ 0 static and fluctuational regimes in KCuF$_3$.  
\begin{figure}
\centering
\includegraphics[totalheight=3.2 in,width=3.4 in,  scale=0.5, angle=0]{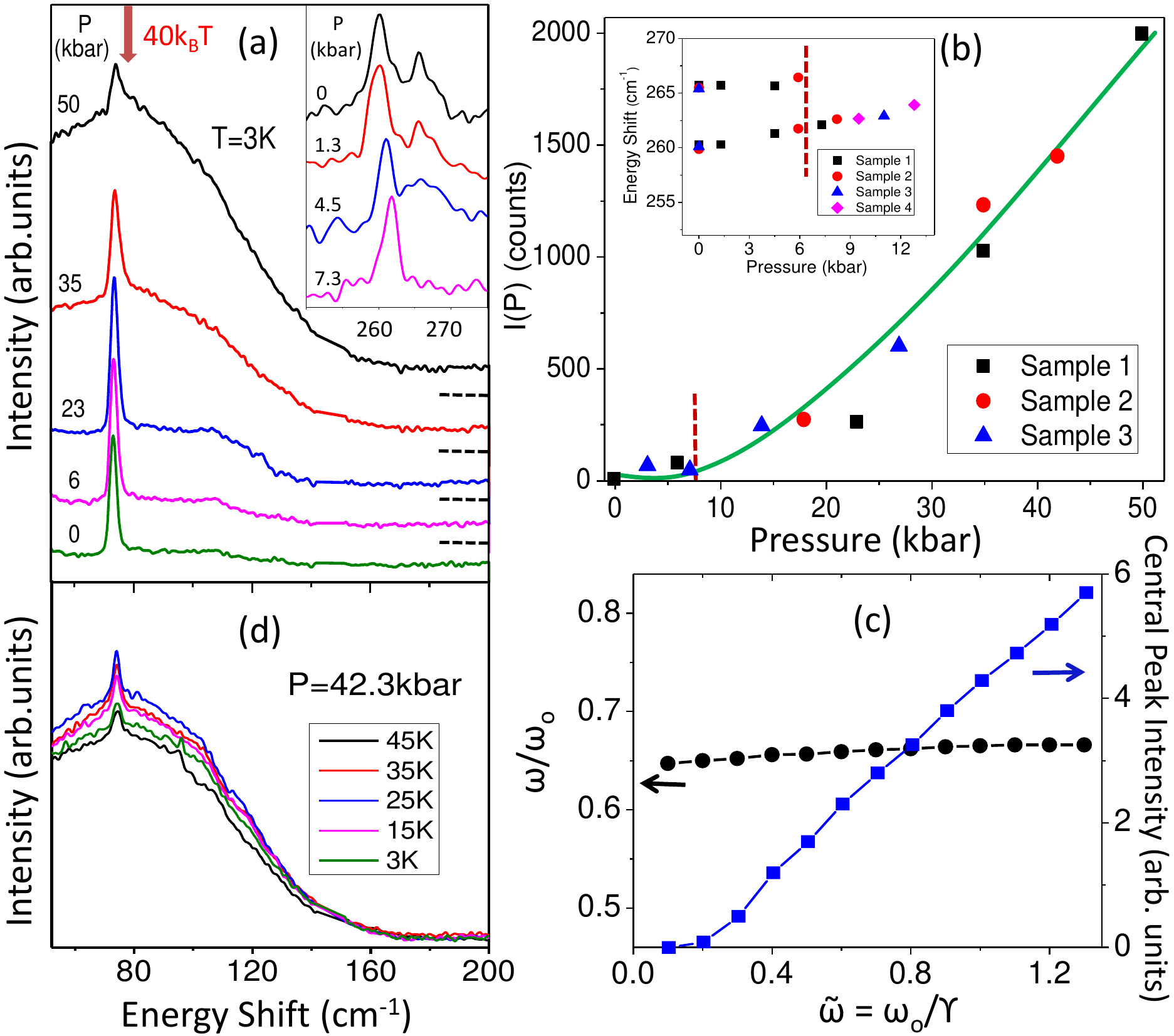}
\caption{(a) Pressure dependence of Raman spectra of KCuF$_3$ at $T$ = 3 K. The arrow indicates a frequency corresponding to 40 $k_{B}T$.  All spectra have the same y-axis scale and have been offset in the y-axis direction for clarity. Dashed lines indicate the common baseline for all the spectra. The inset illustrates the pressure 
dependence of the $E_g$ phonon mode at $T$ = 3 K. (b) Pressure dependence of the integrated quasielastic scattering response intensity, $I(P)$, at $T$ = 3 K for three different samples of KCuF$_3$. The inset shows the pressure 
dependence of the peak energies of the $E_g$ phonon mode at 3 K for four different samples, showing evidence for 
an orthorhombic-to-tetragonal transition near $P^{*}$ = 7 kbar. (c) Calculated normalized phonon frequency, $\omega/\omega_o$, (black circles) 
and quasielastic scattering response integrated intensity (blue squares) as a function of $\omega_{o}/\gamma$, using Eq. 1 from [17]. (d) Temperature dependence of quasielastic scattering response of KCuF$_3$ at $P$ = 42.3 kbar.  All spectra have the same y-axis scale and all spectra have been shifted by the same amount in the -y direction to emphasize the quasielastic contribution to the spectra. The low energy (55 cm$^{-1}$) cutoff in Figs. (a) and (d) reflects the low energy limit of the spectral window defined by our spectrometer.}
\label{fig:test}
\end{figure}

Fig. 2 shows the pressure-dependent Raman spectra of KCuF$_3$. The insets of Fig. 2 (a) and (b) show that the splitting of the $\sim$ 260 cm$^{-1}$ $E_g$ phonon mode disappears 
above $P^{*} \sim$ 7 kbar, revealing a pressure induced orthorhombic-to-tetragonal transition. Figs. 2 (a) and (b) also show that the 
pressure-induced structural transition near  $P^{*} \sim$ 7 kbar (at $T$ = 3 K) is followed by the development with increasing pressure of a 
broad quasielastic response centered at $\omega$=0; this quasielastic scattering response is indicative of fluctuational behavior at low temperatures and high pressures ($P$ $\textgreater$ 7 kbar) in KCuF$_3$, and can be qualitatively described by a simple relaxational response 
function $\chi ^{\prime \prime }(\omega )\sim \frac{\omega \gamma }{\omega ^2+\gamma ^2}$,[25] which has a maximum value at the characteristic fluctuation rate $\gamma$. Because the maximum value in the quasielastic scattering (i.e., $\gamma$) doesn't change appreciably with pressure (see Fig. 2(a)), the increasing quasielastic scattering with pressure in Fig. 2(b) is believed to primarily reflect an increase in the overall amplitude of the quasielastic scattering response, for example due to a systematic increase in the volume of fluctuating regions. Similar fluctuational responses --- albeit with very different characteristic fluctuation rates  ---  have been observed to result from slow relaxational structural fluctuations in SrTiO$_3$ [26], LaAlO$_3$ [27] and KMnF$_3$ [28]. In particular, a fluctuational (diffusive) neutron scattering response in isostructural KMnF$_3$ was also attributed to dynamic rotations of MnF$_6$ octahedra; these octahedral fluctuations were shown to  be highly correlated  --- via the shared F ions --- within the planes, but were shown to fluctuate in an uncorrelated fashion between adjacent planes. [28] Additionally, previous x-ray diffraction studies of KCuF$_3$ [16] show that in-plane correlations between CuF$_6$ octahedra extend no further than $\sim$ 100 unit cells. Consequently, the fluctuational response we observe could involve interplane octahedral fluctuations and/or in-plane fluctuations between correlated regions of order $\sim$ 1000 $\AA$. Pressure-dependent x-ray diffraction measurements are needed to distinguish between these possibilities.    
\begin{figure}
\centering
\includegraphics[totalheight=3 in,width=3.4 in,  scale=0.5, angle=0]{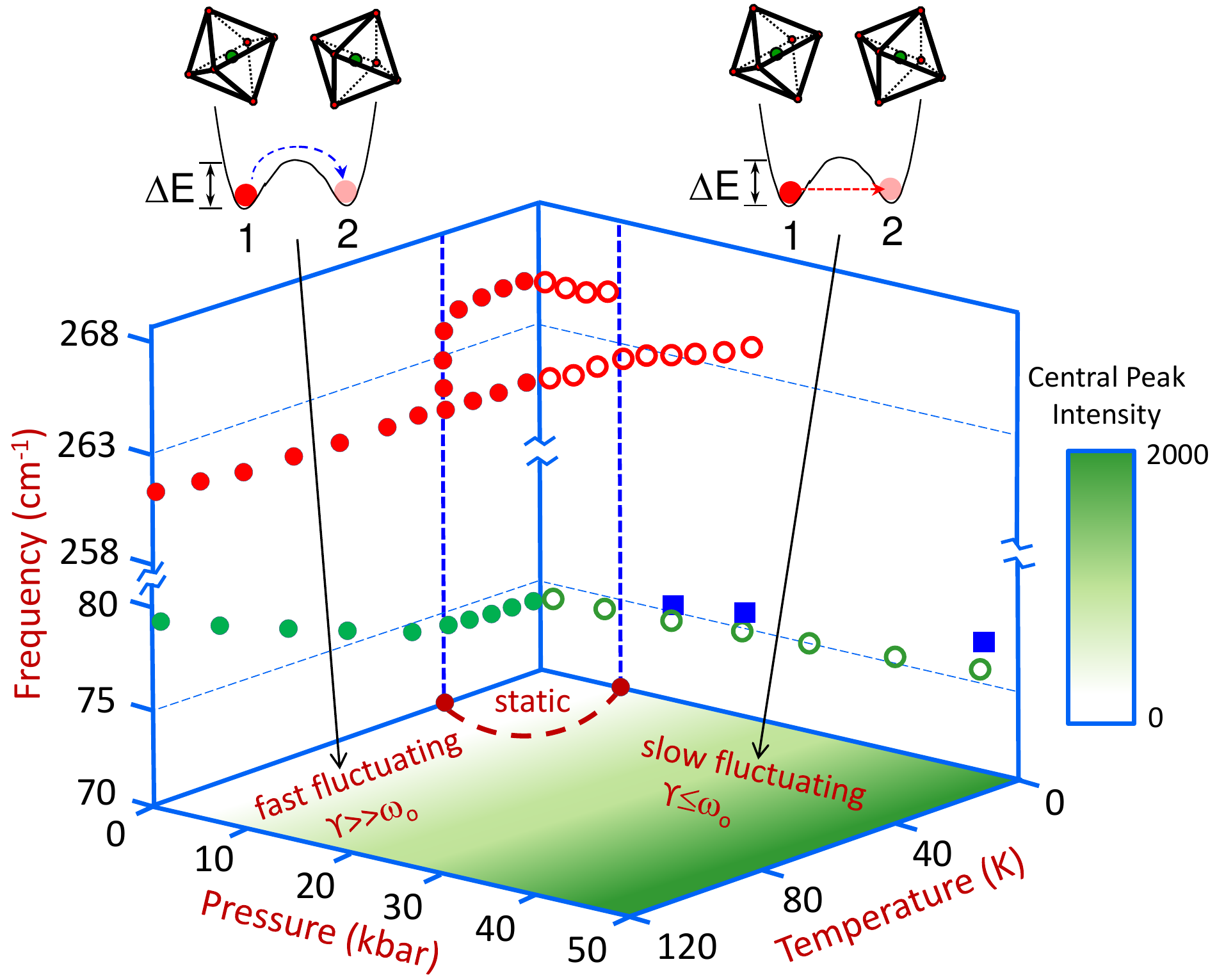}
\caption{$PT$ phase diagram for the CuF$_6$ octahedral fluctuations in KCuF$_3$.  Horizontal axes 
represent the temperature and pressure. The contour plot on the horizontal plane represents the 
measured fluctuational response integrated intensity, with dark green = 2000 counts and white = 0 counts, based on temperature sweeps at the 
following pressures: $P$ = 0, 5, 13, 18.7, 27, 35, 42 kbar. The vertical axis 
shows the mode frequency, with both the $\sim$ 79 cm$^{-1}$ $B_{1g}$ and $\sim$ 261 cm$^{-1}$ $E_g$ phonon frequencies shown as functions of 
temperature (filled red and green circles, respectively) and pressure (open red and green circles, respectively).  Filled squares 
illustrate the characteristic energy $\Gamma$ of the fluctuational response. Diagrams on top depict (left) thermally 
activated hopping between CuF$_6$ configurations in the fast-fluctuating regime of KCuF$_3$, and (right) the quantum 
tunneling between CuF$_6$ configurations in the pressure-tuned slow fluctuating regime.}
\end{figure}

Significantly, all the key spectroscopic features of our temperature- and pressure- dependent Raman results on KCuF$_3$  --- which are summarized in Fig. 3  --- can be qualitatively described by a coupled pseudospin-phonon model [17] in which the normal mode vibrations of a phonon are associated with a molecular group (i.e., the CuF$_6$ octahedra in KCuF$_3$) that fluctuates between discrete configurations and whose dynamics can be described using a pseudospin representation. This coupled pseudospin-phonon model provides a qualitative description of how fluctuations in CuF$_6$ octahedral orientation influence phonon modes (e.g., the $E_g$ and $B_{1g}$ phonons) associated with the fluorine ions in KCuF$_3$. [29] The Hamiltonian for the coupled pseudospin-phonon model is given by, [17]  
\begin{eqnarray}
H&=&\frac{1}{2}\sum_{K} \left\{ P\left( k\right) P^{*}\left( k\right) +\omega
_o^2(k)Q(k)Q^{*}(k)\right\}\nonumber\\
&&-\frac{1}{2}\sum_{i, j}J_{ij}\sigma _i\sigma _j+\sum_{k, j}\frac{\omega _0 (k)}{\sqrt{N}} \textsl{g}(k) Q(k) \sigma _j e^{ik\cdot r_j}\nonumber
\end{eqnarray}

where $Q$ is the normal coordinate of the phonon, $P$ is the conjugate coordinate of $Q$, $\sigma_{i}$ is the pseudospin,  $J_{ij}$
 is the pair interaction between the \textsl{i}th and \textsl{j}th pseudospins, \textsl{g} is the pseudospin-phonon coupling constant, 
and $\omega_o$ is the bare phonon frequency. The identification of the pseudospin with discrete CuF$_6$ octahedral configurations is supported by x-ray diffraction results on KCuF$_3$ showing that discrete CuF$_6$ octahedral orientations lock into a glassy configuration below the structural phase transition.[16]  The coupled phonon response function associated with this Hamiltonian is:[17]
\begin{eqnarray}
\Phi &=&\frac{2\gamma \textsl{k}_{B} \textsl{T} \left( \frac {\textsl{g}}{\gamma \textsl{J}^{\prime }}\right) ^2}{
\left[ \omega ^2-\overline{\omega }_o^2\right] ^2+\omega ^2\Gamma _1^2}
\end{eqnarray}
where $\gamma $ is the pseudospin (CuF$_6$ octahedral orientation) fluctuation rate, $\textsl{J}^{\prime}=\textsl{k}_{B}\textsl{T}-\textsl{J}$ 
is the renormalized exchange coupling, $\overline{\omega }_o\left\{ =\omega _o\left[ 1-\left(\textsl{g}^2/\textsl{J}^{\prime
}\right) \right] ^{1/2}\right\} $ is the renormalized phonon frequency, 
and $\Gamma _1\left\{ =\left( \omega ^2-\omega _o^2\right) /\gamma \textsl{J}^{\prime}\right\} $   is 
the phonon damping parameter.  

The coupled pseudospin-phonon model predicts two regimes of behavior that are qualitatively consistent with the observed pressure- and temperature- dependent Raman results observed in KCuF$_3$:

\textsl{``Soft phonon'' regime}, $\gamma \gg \omega _{o}$  --- When the fluctuation rate ($\gamma$) of the pseudospin (CuF$_6$ octahedral orientation) is much faster than the phonon frequency ($\omega_{o}$), i.e., for $\gamma \gg \omega _{o}$, this model predicts phonon mode softening as the temperature decreases towards the structural phase transition ($T \rightarrow T_c$),[17] as illustrated in the inset of Fig. 1(b) for the case $\gamma = 10 \omega _{o}$. This model prediction is qualitatively consistent with the temperature-dependent mode softening observed for the 50 cm$^{-1}$ $E_g$ (not shown, see [16]) and 72 cm$^{-1}$ $B_{1g}$ (see Figs. 1(a) and (b)) rotational F$^{-}$ phonon 
modes in KCuF$_3$, supporting the conclusion [12, 16] that there is a thermally driven fluctuational regime in KCuF$_3$ in which thermal 
fluctuations of the CuF$_6$ octahedra occur on a faster timescale than the $E_g$ and $B_{1g}$ phonon frequencies to which they are coupled.

\textsl{``Diffusive mode'' regime}, $\gamma \leq \omega _{o}$  --- By contrast, when the fluctuation rate ($\gamma$) of the pseudospin (CuF$_6$ octahedral orientation) is comparable to or slower than the phonon frequency ($\omega_o$), the coupled pseudospin-phonon model (Eq. 1) [17] predicts a ``diffusive mode'' regime, i.e., the development of a $\omega$ = 0 fluctuational response (squares, Fig. 2(c)), and reduced phonon softening (filled circles, Fig. 2(c)). This prediction matches the observed pressure-induced quasielastic response (Fig. 2(a)) and pressure-independent $B_{1g}$ mode frequency (Fig. 2(a) and open green circles, Fig. 3) observed in KCuF$_3$. Thus, the pressure-dependent development of a quasielastic fluctuational response at low temperatures in KCuF$_3$ is consistent with the onset of slow fluctuations (compared to phonon frequencies) of the CuF$_6$ octahedra, which result when the pressure-induced octahedral-to-tetragonal distortion ``unlocks'' the glassy arrangement of CuF$_6$ octahedral tilts.

The pressure results presented here offer evidence for a pressure-tuned ``quantum melting'' transition near $T \sim$ 0 in KCuF$_3$ between a static configuration of the CuF$_6$ octahedra 
to a phase in which the CuF$_6$ octahedra are slowly fluctuating on a timescale that is comparable to or slower than the $E_g$ and $B_{1g}$ phonon frequencies. Because the characteristic rate associated with these CuF$_6$ fluctuations, $\gamma \sim$ 80 cm$^{-1}$ (10 meV), is temperature independent [30] and more than an order-of-magnitude larger than the thermal energies, $\gamma \sim$ 40$k_{B}T$ (arrow in Fig. 2(a)), we propose that these low temperature, pressure-induced fluctuations are primarily driven by zero-point fluctuations (i.e., quantum tunneling) between different wells in the free energy landscape (top right diagram in Fig. 3). This interpretation suggests that the pressure-induced ``quantum melting'' transition in KCuF$_3$ is similar to the ``rotational melting'' transitions [1] to quantum paraelectric phases in SrTiO$_3$ and KTaO$_3$ at low temperatures,[31] and in KH$_2$PO$_4$ (KDP) at high pressures.[32]

One outstanding issue concerns the role these octahedral fluctuations play in disrupting magnetic order in KCuF$_3$. A connection between quantum structural (octahedral) fluctuations and the spin and/or orbital 
degree of freedom might indicate that a pressure-induced orbital/spin liquid state accompanies quantum fluctuations of the 
octahedral orientations in KCuF$_3$. To study this important issue, pressure dependent magnetic measurements are needed 
to test whether the pressure-tuned onset of octahedral fluctuations is coupled with a suppression of N$\acute{e}$el order.  Uniaxial 
pressure measurements would also provide an interesting comparison to these hydrostatic pressure studies,[1] by stabilizing the lower symmetry, static configuration of KCuF$_3$ and thereby favoring the onset of magnetic/orbital order. 

This material is based on work supported by the U.S. Department of Energy, Division of Materials Sciences, under 
Award No. DE-FG02-07ER46453, through the Frederick Seitz Materials Research Laboratory at the University of Illinois at 
Urbana-Champaign, and by the National Science Foundation under Grant NSF DMR 08-56321 (M.K.).  S. L. gratefully acknowledges 
financial support from the DST, Govt. of India through a Ramanujam Fellowship.

%\nocite{*}

%\bibliography{}% Produces the bibliography via BibTeX.

\begin{thebibliography}{0}
\bibitem{25} The simple relaxational response function assumes a frequency-independent relaxational response between two discrete levels. Neither of these assumptions likely applied to the fluctuational response in KCuF$_3$, and therefore detailed fits of this response to the observed data would not provide useful information. While we can't completely rule out the possibility that the response observed in KCuF$_3$ is associated with a standard inelastic Lorentzian response, such a response would imply a strongly overdamped ($\gamma\gg \omega _{o}$) response, and hence the presence of strong fluctuations of the CuF$_6$ octahedra in this pressure and temperature regime. Note also that the thermal Bose factor (1+n($\omega$)) $\sim$ 1 over the temperature and frequency range studied in this paper, so there is a negligible Bose factor contribution in the Raman spectra presented in Figs. 1 and 2.
\bibitem{29} Indeed, neutron scattering studies of isostructural KMnF$_3$ show that the diffuse neutron scattering associated with fluctuations of the MnF$_6$ octahedra are connected with mode softening observed for rotational F modes; see ref. 28.
\end{thebibliography}

\end{document}